\documentstyle[epsfig]{elsart}

\newcommand{\be}{\begin{equation}}
\newcommand{\ee}{\end{equation}}

\def\simlt{\lower.5ex\hbox{\ltsima}}
\def\gtsima{$\; \buildrel > \over \sim \;$}
\def\simgt{\lower.5ex\hbox{\gtsima}}

\def\simlt{\lower.5ex\hbox{\ltsima}}
\def\gtsima{$\; \buildrel > \over \sim \;$}
\def\simgt{\lower.5ex\hbox{\gtsima}}

\def\cm{{\rm\,cm}}

\def\ergcm2{\ {\rm erg~cm^{-2} }}
\def\ergscm2{\ {\rm erg~s^{-1}~cm^{-2} }}

\def\cm2s{\ cm^2 ~s^{-1} }

\def\s{\ifmmode \widetilde \else \~\fi}
\def\={\overline}

\def\spose#1{\hbox to 0pt{#1\hss}}

\def\lta{\mathrel{\spose{\lower 3pt\hbox{$\mathchar"218$}}
     \raise 2.0pt\hbox{$\mathchar"13C$}}}
\def\gta{\mathrel{\spose{\lower 3pt\hbox{$\mathchar"218$}}
     \raise 2.0pt\hbox{$\mathchar"13E$}}}
\def\mincir{\ \raise -2.truept\hbox{\rlap{\hbox{$\sim$}}\raise5.truept  
\hbox{$<$}\ }}                                                          %
\def\magcir{\ \raise -2.truept\hbox{\rlap{\hbox{$\sim$}}\raise5.truept  %
\hbox{$>$}\ }}                                                          %
\def\simlt{\ \raise -2.truept\hbox{\rlap{\hbox{$\sim$}}\raise5.truept   
\hbox{$<$}\ }}                                                          %
\def\simgt{\ \raise -2.truept\hbox{\rlap{\hbox{$\sim$}}\raise5.truept   %
\hbox{$>$}\ }}                                                          %
\def\newline{\par\noindent}

\def\s-z{S-Z}

\begin{document}
\begin{frontmatter}
\title{The non-thermal radiation - Cluster Merger Connection}

\author[Fermi]{Pasquale Blasi} 
\address[Fermi]{NASA/Fermilab Theoretical Astrophysics Group,\\
Fermi National Accelerator Laboratory, Box 500, Batavia, IL 60510-0500}

\begin{abstract}
There is compelling evidence that at least some clusters of galaxies
are powerful sources of non-thermal radiation. In all cases where this
radiation has been detected, a general trend is that high energy densities
of cosmic rays and correspondingly low values of the average magnetic field
are required in order to have a self-consistent picture of the
multiwavelength observations.
Mergers of clusters of galaxies might provide
large enough cosmic ray injection rates and at the same time provide the
mechanism for the heating of the intracluster medium to the observed 
temperatures. In this paper we analyze critically all the components that 
play a role in the non-thermal emission of clusters during a merger, with
special attention to mergers occurred in the past of the cluster. We
outline the consequences of this model for high energy gamma ray observations
and for Faraday rotation measurements of the intracluster magnetic field.
\end{abstract}
\end{frontmatter}

\section{Introduction}

The non-thermal radiation observed in several clusters of galaxies at 
frequencies varying from the radio (see \cite{ens_rev} for a recent review) 
to the UV to the soft and hard 
X-rays, is seriously challenging our understanding. In fact, at least
for clusters with an extended radio halo (for instance Coma) the existence 
of a correspondingly diffuse hard X-ray excess was expected,
simply on the basis of the inverse compton scattering (ICS) of the same
electron population responsible for the radio radiation, the latter being
due to synchrotron emission in the intracluster magnetic field. 
However, after the first 
positive detection of an hard X-ray excess in Coma \cite{fusco}, a 
clear problem arose:
the combination of the synchrotron plus ICS model for the radio and X-ray
emissions suggested a very large cosmic ray energy density and a strength 
of the intracluster magnetic field much smaller that the lower limits 
imposed by Faraday rotation measurements \cite{eilek,tracy}, typically 
larger than a $\mu G$.

Since the main reason for these problems resides in the 
synchrotron plus ICS model, some attempts have been made to look for
alternative interpretations of the hard X-ray excess, in particular invoking
the bremsstrahlung emission from a non-thermal tail in the electron
distribution, produced by stochastic
acceleration \cite{blasi00,ensslin,dogiel}. 
The X-ray spectra can be fitted by this model without
requiring small magnetic fields, although large injection rates of MHD waves 
are needed (compatible with the expectations in a merger of two
clusters of galaxies).

As mentioned above, some clusters also show a UV to soft X-ray excess, which
could also be the result of ICS of low energy electrons, but 
very large cosmic ray energy densities are required in this case as well,
so that some authors are now reconsidering the possibility of a thermal origin
\cite{bonamente01,bonamente02}.

In \cite{blasi99} it was shown that the overall energy density in cosmic rays 
in a cluster is severely constrained by the upper limits on the gamma ray
emission of the cluster, and some new observational tests of 
100 GeV - 1 TeV gamma ray astronomy were proposed, aimed to weigh the
non-thermal content of clusters of galaxies. 

A serious effort to find powerful sources of cosmic rays in clusters
was done in \cite{bbp} (see also references therein): ordinary galaxies, 
radio galaxies, accretion shocks
and a possible bright phase in the galaxy evolution were investigated, 
but the resulting energy densities found there fall short of the
ones required to explain the multiwavelength observations by a factor 10-100.
An analysis of the energetic requirements for the radiating electrons in
clusters was presented in \cite{ens98} while the corresponding proton 
component was studied in detail in \cite{voelk,bbp,ens97}. 
More recently an increasing interest has
been shown for mergers of clusters of galaxies as events responsible for 
the heating of the cluster, through the formation and propagation of strong 
shocks in the intracluster medium (see for instance \cite{sar_proc} and
references therein). 
The basic question that we want to address in this paper is whether these
merger related shocks can accelerate particles by the first order Fermi 
mechanism (for a review on Fermi acceleration see \cite{BE}), to the level 
required to explain the 
observations at different wavelengths without invoking anything but 
the synchrotron emission and ICS and without violating the gamma ray limits,
when available.
Since most of the previous calculations concentrated either on the primary 
electrons (neglecting electrons produced as secondaries of cosmic ray 
interactions) or on the secondary electrons (without considering the electrons 
directly accelerated at the shocks), in this paper we will try to include all
these components in a self-consistent way. 

The aim of this paper is to draw some general conclusions, as much independent
as possible of a specific cluster, though we also apply our results to the
case of Coma, as an example.

The paper is structured as follows: in section 2, we describe our current
understanding of the merger between two clusters of galaxies; in section 3
we describe our approach to the transport of the electronic and hadronic
components; in section 4 we summarize the radiation process relevant for
the production of radio, X, UV and gamma rays. In section 5 
we study the conditions that result in the observed
non-thermal phenomena; in section 6 we apply our calculations to the case of
the Coma cluster. Our discussion and conclusions are presented in 
section 7.

\section{Clusters mergers: thermal heating and particle acceleration}

Mergers between clusters of galaxies are among the most energetic phenomena
in the universe. During the merger event the formation of strong shocks is
practically unavoidable and a substantial fraction of the gravitational
energy is released in the form of heating of the intracluster medium. 
The overall energy in the event can be estimated as
\begin{equation} 
E_{merger}\sim \frac{G M_1 M_2}{d} \sim 1.4\times 10^{64} 
\left(\frac{M}{5\times 10^{14} M_{\odot}}\right)^2
\left(\frac{d}{1.5 Mpc}\right)^{-1} {\rm ergs},
\label{eq:lumi}
\end{equation}
where on the left-hand side $M_{1,2}$ represent the total masses of the two 
clusters, and on the right-hand side they have been normalized to the same
value of $5\times 10^{14} M_{\odot}$, typical of average clusters. 
$d$ is a
fiducial value for the spatial scale where the dissipation through
collisionless shocks is likely to occur, taken here to be $\sim 1.5$ Mpc. 
It is easy to verify that $E_{merger}$ is roughly 
comparable with the total thermal energy in the intracluster medium (ICM)
of a rich cluster, which justifies the idea that mergers may play a role 
in the heating of the ICM.

The process of heating of the ICM during merger events has been studied in 
detail in numerical simulations (e.g. \cite{rsm,rsb,paul,KK,rlb}). These 
studies suggest that the 
main channel of energy dissipation during a merger is associated with the
heating of the gas, leaving a few tens of percent available for additional
processes as particle acceleration to suprathermal energies. This is 
consistent with the usual picture of shock acceleration at supernova remnants,
where $1-10\%$ of the kinetic energy of the plasma is transformed into 
energy of non-thermal particles.
The numerical simulations also provide valuable information about the fluid
dynamics of these shocks: for instance typical velocities of 
$v\simgt 10^8$ cm/s
and typical compression factors $r\approx 3-4$ are achieved
\cite{naito}. The typical 
duration of a merger event is therefore $t_{merger}\sim d/v \sim 10^9$ yrs.
Consequently the total rate of energy release during the merger is 
$L_{merger}\sim E_{merger}/t_{merger}\sim 4\times 10^{47} 
\left(\frac{M}{5\times 10^{14} M_{\odot}}\right)^2
\left(\frac{d}{1.5 Mpc}\right)^{-1}$ erg/s. 

Some fraction of this energy (left as a parameter) will be converted to 
kinetic energy of suprathermal particles, due to Fermi acceleration. In the
following we estimate the maximum energy of electrons and protons accelerated
at the merger shocks and their spectrum. 
The acceleration time can be written as follows:
\begin{equation}
\tau_{acc} (E) = \frac{3}{u_1-u_2} D(E) \left[\frac{1}{u_1}+\frac{1}{u_2}
\right] = \frac{3 D(E)}{v^2} \frac{r(r+1)}{r-1},
\end{equation}
valid for any choice of the diffusion coefficient. Here $u_1$ and $u_2$ 
are the upstream and downstream velocities of the fluid and $r=u_1/u_2$
is the compression factor.

We consider two possibilities for the diffusion coefficient $D(E)$. First 
we use the expression proposed in \cite{blasicola99}: 
\begin{equation} 
D(E) = 2.3\times 10^{29} B_{\mu}^{-1/3} L_{20}^{2/3} E(GeV)^{1/3} cm^2/s,
\label{eq:dif1}
\end{equation}
where $B_{\mu}$ is the magnetic field in microgauss and $L_{20}$ is the 
largest scale in the magnetic field power spectrum in units of 20 kpc. 
Here we assumed that the magnetic field is described by a Kolmogorov power 
spectrum.

In this case the acceleration time becomes:
\begin{equation}
\tau_{acc} (E) \approx 6.9\times 10^{13} B_{\mu}^{-1/3} L_{20}^{2/3} 
E(GeV)^{1/3} v_8^{-2} g(r)~~~ s,
\end{equation}
where $v_8=\frac{v}{10^8 cm/s}$ and $g(r)=r(r+1)/(r-1)$ and $v=u_1$.

For electrons, if the average magnetic field is less than $\sim 3\mu G$, the
energy losses are dominated by ICS off the microwave background, with a loss
time $\tau_{loss}\approx 4\times 10^{16}/E~s$, where $E$ is in GeV. The maximum
energy of accelerated electrons is obtained requiring $\tau_{acc}<\tau_{loss}$:
\begin{equation}
E_{max}^e \approx 118 L_{20}^{-1/2} B_{\mu}^{1/4}  
v_8^{3/2} g(r)^{-3/4}~ GeV,
\end{equation}
The compression ratio $r$ and the velocity $v_8$ are not independent, since 
\begin{equation}
r=\frac{\frac{8}{3} {\cal M}^2}{\frac{2}{3} {\cal M}^2 + 2},
\end{equation}
valid for an ideal monoatomic gas. Here ${\cal M}$ is the Mach number of
the unshocked gas, moving with speed $v_8$. 

For protons, energy losses are not relevant and the maximum energy is clearly
determined by the finite time duration of the merger event. Therefore the 
maximum energy for protons will be defined by the condition $\tau_{acc}<
t_{merger}$, which gives 
\begin{equation}
E_{max}^p \approx 9\times 10^7 L_{20}^{-2} B_{\mu} v_8^6 g(r)^{-1/2}~GeV.  
\end{equation}

The second possibility for the diffusion coefficient is a Bohm diffusion,
well motivated for the case of strong turbulence. In this case:
\begin{equation}
D(E)=3.3\times 10^{22} E/B_\mu ~ cm^2/s.
\label{eq:dif2}
\end{equation}
This represents the smallest diffusion coefficient (although it can 
be further decreased in the case of perpendicular shocks \cite{jok}) 
and implies
considerably larger maximum energies than the ones estimated above.
For electrons we obtain:
\begin{equation}
E_{max}^e \approx 6.3\times 10^4 B_{\mu}^{1/2} v_8 g(r)^{-1/2}~ GeV,
\end{equation}
while for protons
\begin{equation}
E_{max}^p \approx 3\times 10^9 B_{\mu} v_8^2 g(r)^{-1}~GeV.  
\end{equation}
If $E_{max}^p$ becomes larger than $\sim 10^{10}$ GeV energy losses due
to pair production and photopion production on the photons of the microwave
background become important and limit the maximum energy to less that 
a few $10^{10}$ GeV. 

Some final comments on the spectrum of accelerated particles are in order. 
Standard shock acceleration theory predicts that the suprathermal particles 
generated by the Fermi process have a power law spectrum in momentum, 
$Q(p)dp\propto p^{-\sigma} dp$ where $\sigma=(r+2)/(r-1)$. As mentioned above, 
numerical simulations predict compression ratios in the range $r=3-4$, which
correspond to $\sigma=2-2.5$ \cite{naito}. 

The classical dilemma of normalizing the electrons and protons spectra exists
here as well as for the much better studied case of supernova remnants. It 
is usually assumed (motivated by observations of the relative abundance
in the Galaxy) that at $E\sim 1$ GeV the protons overcome the electron
number density by a factor 10-100. More conservatively we introduce a parameter
$\xi<1$ representing the ratio of the injection spectra of electrons and
protons (this fraction is later changed by propagation effects). 

Theoretically,
the motivations for having a small $e/p$ ratio are in the microscopic 
processes responsible for the acceleration: protons resonate with Alfv\'en
waves on a very wide range of momenta, and it is therefore not too difficult
to extract them from the thermal distribution and inject them into the
acceleration region. For electrons this is much harder. Low energy electrons,
close to the thermal distribution, do not interact with Alfv\'en waves and 
some other modes need to be excited (for instance whistlers) and
sustained against the strong damping. If these wave modes are excited they 
might be responsible for the injection of a fraction of the thermal electrons
into the acceleration region. A more detailed discussion of the electron
injection at non relativistic shocks can be found in \cite{drury,levinson}
and references therein.

There is a even more general reason for the minor efficiency of electron
acceleration: a particle is accelerated at the shock only if it can ``feel''
the shock, which means that its Larmor radius must be larger than the 
thickness of the shock. The latter quantity is determined by the Larmor
radius of thermal protons (see \cite{Bell1,Bell2} and references therein), 
therefore,
for a proton temperature of $\sim 8$ keV, only electrons of energy larger than 
$\sim 5-10$ MeV can be injected in the acceleration box 
(note that this value is much larger than the typical electron temperature
in the ICM). In the following we will
adopt the value of $5$ MeV as a low energy cutoff in the injection spectrum 
of electrons.
Although the  details of the electron injection are still not completely 
clear, it seems quite plausible that shock acceleration
works more efficiently for the proton component rather than for the electrons.
It is extremely important to keep in mind a fundamental difference between 
electrons and protons in clusters of galaxies: while electrons suffer 
severe energy losses and they can only trace the recent activity of the
cluster, protons are stored on cosmological time scales \cite{bbp,voelk} 
without
appreciable losses, being therefore able to produce secondary products
(electrons, gamma rays and neutrinos) at any time.
It is therefore of great importance to include the hadronic component in any
calculation of non-thermal phenomena in clusters of galaxies.

\section{The transport of non-thermal particles}

The passage of the shock through the cluster heats the gas and possibly
accelerates a fraction of electrons and protons to suprathermal energies.
The injection spectra are modified by the propagation (diffusion) in the
intracluster magnetic field and by energy losses, therefore the electron
and proton spectra at each time must be calculated by solving a transport 
equation, in the form:
$$
\frac{\partial n_j(E_j,r,t)}{\partial t} = 
$$
\begin{equation}
\frac{1}{r^2} 
\frac{\partial}{\partial r} \left[r^2 D_j(r,E_j) 
\frac{\partial n(E_j,r,t)}{\partial r}\right]
+ \frac{\partial}{\partial E_j} \left[b_j(r,E_j) 
n_j(E_j,r,t)\right]
+ Q_j(E_j,r,t)~, 
\label{eq:transport}
\end{equation}
where the index $j$ labels protons and electrons, $D$ is the diffusion
coefficient, in general dependent on energy and distance, $b_j$ is the 
total rate of energy losses, $Q_j$ is the rate of injection of particles,
and finally $n_j$ is the spectrum of the $j-$th
component resulting from diffusion and losses. We are assuming here that 
the cluster has a spherical symmetry, so that the spatial dependence is
all contained in the radial coordinate $r$.
For electrons, the Coulomb, bremsstrahlung, synchrotron and inverse
Compton scattering energy losses have been included, while for protons
the only relevant loss channel is the $pp$ scattering.

A discussion of the role of the diffusion coefficient is in order: in
section 2, two possibilities were considered and the results for the 
maximum energy of accelerated electrons and protons were very different, as
expected. Such large uncertainties in the diffusion coefficient do not 
affect the solution of eq. (\ref{eq:transport}). The time evolution of the
electron component (both primary and secondary) is largely dominated by 
the fast energy losses, so that the use of eq. (\ref{eq:dif1}) or 
of eq. (\ref{eq:dif2}) for the diffusion coefficient makes no difference
(in other words, the first term on the left hand side of eq. 
(\ref{eq:transport}) is small). For the proton component the situation is
different, because energy losses are very weak. However, the diffusion times
out of the cluster are large, and indeed larger than the age of the cluster,
for the energies that we are interested in. This phenomenon of confinement
of cosmic rays, was widely discussed in \cite{bbp}, where
the concept of clusters of galaxies as ``storage rooms'' for cosmic rays
was first introduced.
It was showed there that the spectra of secondaries (electrons, gamma rays, 
neutrinos)
depend only on the properties of the injection spectrum of cosmic rays,
and are not affected by the details of the diffusion. 

The injection of electrons at the shock is assumed to be a power law in
momentum $Q_e(p)=f(r) p^{-\sigma}$, where the function $f(r)$ is taken to be
proportional to the local density profile in the cluster, which can be written
as:
\begin{equation}
n_{gas}(r) = \frac{n_0}{\left[1+\left(\frac{r}{r_c}\right)^2\right]^{\frac
{3}{2}\beta}}.
\end{equation}
Here $n_0$ is the density in the inner core, $r_c$ is the core radius, and 
$\beta$
is a phenomenological parameter (see \cite{sarazin} for a review). 
The normalization constant in the injection spectrum of electrons, $Q_e(p)$, 
is chosen in order to fit the data.

For the protons the injection spectrum is also a power law in momentum with
the same radial dependence, but the absolute normalization is parametrized
in terms of the ratio $\xi$ of the injection rate of electrons and protons.

Once the spectrum of protons is calculated by solving eq. (\ref{eq:transport})
for the proton component, the injection spectra of secondary electrons 
generated by the decay of charged pions is completely determined. For 
secondary electrons the injection rate $Q_e$ is 
$$
Q_e(E_e,r)= \frac{m_{\pi}^2}{m_{\pi}^2-m_{\mu}^2} n_{gas}(r) c \cdot
$$
\begin{equation}
\int_{E_e}^{E_p^{max}} dE_\mu \int_{{E_\pi}^{min}}^{{E_\pi}^{max}}
\frac{dE_\pi}{E_\pi} \int_{E_{th}(E_\pi)}^{E_p^{max}} dE_p~G_\pi(E_\pi,E_p)
F_e(E_e,E_\mu,E_\pi) n_p(E_p,r) ~,
\label{eq:source}
\end{equation}
where $E_p^{max}$ is the maximum proton energy (note that 
our calculations are insensitive to the exact value of $E_p^{max}$),
$E_{th}(E_\pi)$ is the threshold energy for the production of pions with
energy $E_\pi$ and we put
$$
E_\pi^{min}=\frac{2 E_\mu}{(1+\beta)+\delta (1-\beta)}~;~~~
E_\pi^{max}=min\left\{E_p^{max},
\frac{2 E_\mu}{(1-\beta)+\delta (1+\beta)}
\right\}~,
$$
where $\delta=m_\pi^2/m_\mu^2$ and $\beta$ is the velocity of muons in units 
of the light speed.
The quantity $n_p(E_p,r)$ is the CR spectrum at distance
$r$ from the source, as given by eq. (\ref{eq:transport}).

The interactions are described by the functions
$G_\pi(E_\pi,E_p)$ (the differential cross section for pions with energy 
$E_\pi$ produced in a CR interaction
at energy $E_p$ in the laboratory frame) and $F_e(E_e,E_\mu,E_\pi)$
(the spectrum of electrons in the decay of a muon of energy $E_\mu$
produced in the decay of a pion with energy $E_\pi$). The electron
spectrum is given by the following expression:
$$
F_e(E_e,E_\mu,E_\pi)=\frac{1}{\beta E_\mu}\times
$$
\begin{equation}
\times 
\left\{
        \begin{array}{l l}
                2\left(\frac{5}{6}-\frac{3}{2}\lambda^2+\frac{2}{3}
 \lambda^3\right)-P_\mu\frac{2}{\beta}\left[\frac{1}{6}-
\left(\beta+\frac{1}{2}\right)\lambda^2+\left(\beta+\frac{1}{3}\right)
\lambda^3\right]~~~~~ &
              \mbox{if $\frac{1-\beta}{1+\beta}\leq \lambda\leq 1$},\\

       \frac{4\lambda^2\beta}{(1-\beta)^2}\left[3-\frac{2}{3}\lambda
	\left(\frac{3+\beta^2}{1-\beta}\right)\right]-\\
\frac{4P_\mu}{1-\beta}\left\{\lambda^2(1+\beta)-\left[
\frac{1}{2}+\lambda (1+\beta)\right] \frac{2\lambda^2}{1-\beta}+
\frac{2\lambda^3 (\beta^2+3)}{3(1-\beta)^2}\right\} &
                     \mbox{if $0\leq\lambda\leq \frac{1-\beta}{1+\beta}$},
        \end{array}
        \right\}
\end{equation}
where we put
\begin{equation}
P_\mu=P_\mu(E_\pi,E_\mu)=\frac{1}{\beta}\left[
\frac{2E_\pi \delta}{E_\mu(1-\delta)}-\frac{1+\delta}{1-\delta}\right],
\end{equation}
and $\lambda=E_e/E_\mu$.
The above expression for $F_e$ takes into account that the
muons produced from the decay of pions are fully polarized (this
is the reason why the pion energy $E_\pi$ appears in the expression for
the electron spectrum from the muon decay).

Determining the pion distribution in the low energy region (pion energies 
close to the mass of the pions) is not trivial. A satisfactory
approach to the low energy pion production was proposed in
\cite{stecker,dermer} and recently reviewed in \cite{strong} in the context
of the {\it isobaric model}. The detailed and lengthy expressions
for $G_\pi$ that we used are reported and discussed in details in
\cite{strong} (see their Appendix). Thus, following \cite{strong}, 
we use here their
model for collisions at $E_p\lta 3$ GeV. For $E_p\gta 7$ GeV we use
the scaling approximation which can be formalized writing the
differential cross section for pion production as
\begin{equation}
d \sigma/ d E_{\pi} = (\sigma_0/E_{\pi}) f_{\pi^\pm}(x),
\label{eq:scal}
\end{equation}
where $\sigma_0=3.2\cdot 10^{-26}~cm^2$, $x=E_{\pi}/E_p$.
The scaling function $f_{\pi^\pm}(x)$ is given by
\begin{equation}
f_{\pi^\pm}(x)=1.34 (1-x)^{3.5} + e^{-18x}.
\label{eq:fpi}
\end{equation}
In this case the function $G_\pi$ coincides with the definition of
differential cross section given in eq. (\ref{eq:scal}).

From the discussion above, it is clear that in the case of secondary 
electrons the transport equations for protons and secondary electrons 
are coupled. 

In the following sections we will solve the transport equations for the
different components using some {\it fiducial} values for the basic
parameters for the cluster merger, and for different possibilities for the time
evolution of the merger event. 

\section{Non-thermal radiations}

In this section we briefly summarize the mathematical basis for the
calculation of the radio, UV, X-ray and gamma ray spectra. Four processes
are involved in the production of the radiation: {\it a)} synchrotron 
emission; {\it b)} inverse Compton scattering, {\it c)} pion 
production and decay, and {\it d)} bremsstrahlung of primary and secondary 
electrons. We discuss them separately.

\subsection{Synchrotron emission and radio radiation}

High energy electrons produce radio radiation in $\sim \mu G$
magnetic fields by synchrotron emission. The energy generated per
unit time, per unit volume, per unit frequency $\nu$ is given by \cite{sar}
\begin{equation}
{\cal L}(\nu) = \frac {3^{1/2} e^3 B}{m_e c^2} \int n(\gamma,r,t) 
{\cal R}(x) d\gamma,
\end{equation}
where $e$ and $m_e$ are the electron charge and mass respectively, $B$ is 
the magnetic field, $c$ is the speed of light and $n_e$ is the solution 
of the transport equation for electrons (primaries or secondaries) with 
kinetic energy $m_e c^2 (\gamma-1)$ at the position at distance $r$ from
the cluster's center at the time $t$. The function ${\cal R}(x)$ is defined
as
\begin{equation}
{\cal R}(x) = 2 x^2 \left\{K_{4/3} (x) K_{1/3}(x) - \frac{3}{5} x 
\left[ K_{4/3}^2(x) - K_{1/3}^2(x)\right]\right\}
\end{equation}
and the $K_y$'s are modified Bessel functions. The variable $x$ is function of
$\gamma$ through $x=\nu/(3\gamma^2 \nu_c)$, where the critical frequency is
$\nu_c=eB/(2\pi m_e c)$.

\subsection{Inverse Compton Scattering (ICS)}

The electrons of relevance in clusters of galaxies produce photons by 
ICS off the CMB background in a wide range of accessible frequencies, from 
UV to soft X-rays to hard X-rays and gamma rays. 

The particle number density per unit volume, per unit time, per unit
photon energy $E_{ph}$ at distance $r$ from the cluster's center is given by 
\begin{equation}
{\cal Q}_{ph}(E_{ph}) = \frac{12\pi\sigma_T}{h E_{ph}} \int_{1}^\infty
d\gamma n_e(\gamma,r,t) \int_0^1 dx {\cal G}(x) {\cal J}
\left(\frac{E_{ph}}{4\gamma^2 h x}\right),
\end{equation}
where $\sigma_T$ is the Thomson cross section, ${\cal G}(x)=1+x+2x \ln x -2
x^2$ and
\begin{equation}
{\cal J} (\tilde \nu) = \frac{2 h {\tilde \nu}^3}{c^2}
\frac{1}{exp(h{\tilde \nu}/(kT_{CMB})) - 1}
\end{equation}
is the brightness of the CMB background at temperature $T_{CMB}$.
 
\subsection{Neutral pion production and decay}

Gamma rays are produced by the decay of neutral pions ($\pi^0 \to
\gamma\gamma$) generated in $pp$ inelastic interactions. This channel 
usually provides an important contribution to the gamma ray fluxes 
above $\sim 100$ MeV.

The emissivity in gamma rays at distance $r$ and
energy $E_\gamma$ is given by
\begin{equation}
j_\gamma^{\pi^0}(E_\gamma,r)=2 n_H(r) c
\int_{E_\pi^{min}(E_\gamma)}^{E_p^{max}}
dE_\pi \int_{E_{th}(E_\pi)}^{E_p^{max}}
dE_p G_{\pi^0}(E_\pi,E_p)\frac{n_p(E_p,r)}{(E_\pi^2-m_\pi^2)^{1/2}},
\label{eq:gamma1}
\end{equation}
where $E_\pi^{min}(E_\gamma)=E_\gamma+m_{\pi^0}^2/(4E_\gamma)$.
We refer to \cite{strong} for the expression of $G_{\pi^0}(E_\pi,E_p)$
in the low energy collisions ($E\leq 3$ GeV), while we use 
the scaling approach
given in eqs. (\ref{eq:scal}) and (\ref{eq:fpi}) for $E_p>7$ GeV,
with $f_{\pi^0}=(1/2)f_{\pi^\pm}$.\par

\subsection{Bremsstrahlung}

The flux of gamma rays due to bremsstrahlung of primary and secondary electrons
is given by
\begin{equation}
j_\gamma^{brem}(E_\gamma,r)=n_{gas}(r) c\int_{E_\gamma}^{E_p^{max}}
dE_e n_e(E_e,r) \frac{d\sigma}{dE_\gamma}(E_e,E_\gamma),
\end{equation}
where the differential cross section can be approximated as
\begin{equation}
\frac{d\sigma}{dE_\gamma}(E_e,E_\gamma)=2.6\cdot 10^{-26}/E_\gamma.
\end{equation}

\section{The non-thermal side of mergers}

Mergers are able to heat the ICM to the observed temperature through
the formation of collisionless shocks. In the same process, some fraction
of the thermal particles acquires a suprathermal energy distribution,
by first order Fermi acceleration at the same shocks, so that relativistic
particles are likely to be generated during the merger. 
In this section we study the non-thermal appearance of a cluster for different
scenarios of a merger. 

Since it is not our purpose to study a specific cluster in detail, we will 
choose a set of fiducial values for the parameters and use those values in
our calculations in order to derive some general trends. The dependence of the
values of the parameters will be discussed. The
total merger luminosity is taken from eq. (\ref{eq:lumi}), the magnetic field 
is taken to be $B=0.15\mu G$, but the effect of higher fields is studied too.
We emphasize the fact that observations force to use these low fields if the
X-ray emission is interpreted as ICS. Therefore one must either believe
the results of Faraday rotation measurements (larger fields) and look for 
alternative interpretation of the hard X-ray excess or have a good reason 
to discard the results of Faraday rotation measurements and maintain an ICS
interpretation for the hard X-ray excess.
Indeed it is quite simple to understand this effect: 
both the radio and ICS fluxes depend linearly on the energy injection rate
in the form of electrons, but the former scales with the magnetic field 
as $B^{1+\gamma/2}$, where 
$\gamma$ is the power index of the electron injection spectrum. Therefore, 
at fixed X-ray (ICS) flux, the radio flux increases very fast with increasing
$B$. For instance for a magnetic field of $1\mu G$ the radio flux would 
increase by a factor $\sim 60$ in comparison to the case $B=0.15\mu G$
(we use here $\gamma=2.32$).
This can be seen in another way: if the radio flux is explained in terms of
a $\mu G$ magnetic field, then the X-ray (ICS) flux will be $\sim 60$ times
smaller than for the case $B=0.15\mu G$. If $B=5\mu G$ the factor $60$ becomes
$\sim 2000$. 

For clusters which are experiencing a current merger, the non-thermal activity
strongly depends on the stage at which the merger is observed. In these cases
there are usually strong signatures that particle acceleration is occurring
(see the case of A3667 \cite{msv}), and they could be used as a testing 
facility for
the ideas presented in this paper. However they represent a subsample of
the clusters which show non-thermal activity.

The interesting and most likely case is that of a merger occurred in
the past, since most or all of the clusters are supposed to have had 
at least one merger during their lifetime.
It is easy to understand that the results will depend on the
time from the merger, when compared with the cooling time of high energy 
electrons. After $\sim 10^9$ yrs the high energy electrons responsible for the
radio and hard X-ray emissions will have cooled through ICS and synchrotron
emission, so that the observed spectra will have no radio and hard X-ray
emission or have a very weak one. On the other hand electrons will have piled
up at low energy, providing a bulk of potentially powerful UV and soft X-ray
emitters. Cluster mergers which occurred $\sim 10^9$ years ago will be on the
boundary and might or might not show high energy non-thermal emission. These
conclusions are based on the primary electrons only, and have also been 
derived in some different form in \cite{sar}. A crucial question is
what happens to the hadronic component. In fact, protons do not lose energy in
appreciable way on the time scales we are interested in 
\cite{bbp,cb,blasicola99}. 
Therefore, even if they have been injected in the cluster during a merger
in the very past, they are still efficient in generating secondary electrons 
and high energy radiation at present. 

This is a very solid conclusion: if all clusters have experienced a powerful 
merger at some point of their lifetime, then all of them should have to some
extent 
non-thermal activity, because the proton-induced electron spectra cannot be 
depleted by energy losses. In other words the secondary electron injection
is continuous even if the proton injection occurred in the past of the 
cluster. The detectability of this radiation depends on conditions which 
are specific of the single clusters.

Here we formally calculate the contribution of primary and secondary electrons
for different times elapsed from the last merger and show more rigorously what
stated above. 

The technique used in the calculations is the one illustrated in the previous
sections: we inject an electron and proton component, both having a spectrum 
modelled as a power law in momentum and such that the ratio of the two spectra
at injection  is given by the parameter $\xi$ (since the spectra are both power
laws in momentum this ratio is kept at all momenta).
Some comments are required about the total energy content of
the electron and proton components: the total energy density injected per unit
time in the component $i$ ($i=$protons, electrons) is 
\begin{equation}
E_i = \int_{p_{min}} dp q_0 p^{-\gamma} \epsilon(p),
\end{equation}
where we wrote both electron and proton spectra in the form 
$q_{0,e/p} p_{e/p}^{-\gamma}$.
For the purpose of an order of magnitude estimate let us assume that the 
energy $\epsilon(p)$ of a particle with momentum $p$ can be approximated 
by $p^2/(2m)$ for $p<m$ and by $p$ for $p>m$. In this case it is easy to
show that the ratio of energy injections is
\begin{equation}
\frac{E_e}{E_p} \approx \frac{2\xi(3-\gamma)}{4-\gamma} 
\left(\frac{p_e^{min}}{m_p}\right)^{-\gamma+2}.
\end{equation}
For the value $\gamma=2.32$ that we use here (correspondent to 
$r=3.27$,
compatible with the values found in numerical simulations), and for 
$p_e^{min}\approx 5$ MeV,
obtained above, we obtain $E_e/E_p\approx 4.3 \xi$. In other words, if 
the number density in protons is larger at fixed $p$ by a factor $1/\xi=10$ 
(100), the correspondent energy injection rate in protons is a factor 
2.3 (23) larger than the energy injection rate in electrons. 
The correspondent ratio of the rate of injections of particles is 
$Q_e/Q_p\approx \xi (p_e^{min}/p_p^{min})^{-\gamma+1}$. If we take 
the minimum kinetic energy of the protons to be $\sim 20$ keV,
and $\gamma=2.32$ we have $Q_e/Q_p\approx 1.3 \xi=0.13$ (0.013) 
for $\xi=0.1$ (0.01).

Both injections are assumed to occur in a way proportional to the local 
gas density in the ICM. There is a theoretical justification for this: if the
only discrimination between a particle that can be accelerated and one that
cannot is in the particle energy, and if we assume that the temperature of the
cluster is the same at all points, then the fraction of particles energized 
at the shock must be simply proportional to the local gas density. This simple
picture might be modified because the injection of particles in the
acceleration region is thought to occur through resonant scattering with waves,
and these waves might be damped more efficiently in the denser regions of
the cluster.

For the protons, the rate of secondary electrons
is consistently calculated and used in the transport equation to calculate 
their effective local spectrum at each time. A very important point must be
noticed: after the proton injection has stopped (the merger is over) the 
rate of production of secondary electrons does not change with time, because
diffusion is too slow to allow the proton escape from the cluster, and the
energy losses of protons do not affect their spectrum. By all means
the protons work as a continuous source of ``new'' electrons in the cluster,
even if the process that generated them is no longer operating. 

Our results are summarized in Figs. 1 and 2 showing the radio emission and
the UV plus X-ray emission of the cluster, for a typical distance of
100 Mpc (the results can be easily rescaled to an arbitrary distance). 
In Fig. 1 the dashed thin
lines represent the radio emission from the primary electron component 
$0,~ 5\times 10^8$ and $10^9$ yrs after the injection of new particles
is finished (as indicated on the plot). These lines are obtained by adopting a
diffusion coefficient as in eq. (\ref{eq:dif1}) with $v_8\approx 4.8$ and 
$L_{20}=25$ (500 kpc).
The thick dashed line represents the radio flux from primary electrons at 
$t=0$ in the case of Bohm diffusion [eq. (\ref{eq:dif2})]. It can be clearly
seen that only the tail of the radio spectrum is affected by the choice of a
different diffusion coefficient (because in this 
second case the maximum energy of the primary electrons is larger) while the
lower frequency spectrum remains almost unaltered (within $10-20\%$), although
the two diffusion coefficients are very different. For the curves describing
the radio emission at later times, there is no difference as a result of a 
different choice of the diffusion coefficient. 
The solid lines represent the
radio emission from secondary electrons for the indicated values of $\xi$.
Due to confinement of the primary cosmic rays, these curves are not affected
by the choice of the diffusion picture \cite{bbp}. All the curves mentioned
above are obtained for a magnetic field of $0.15\mu G$, but it is important to
show the effect of larger fields. As a reference, we plotted as a dash-dotted
curve the radio flux for a magnetic field of $1\mu G$. We find that the rate
of injection of electrons must be $\sim 60$ times smaller than before to
generate the same radio emission as for the case $B=0.15\mu G$. This 
has important consequences for the X-ray fluxes (see below).

\begin{figure}[thb]
 \begin{center}
  \mbox{\epsfig{file=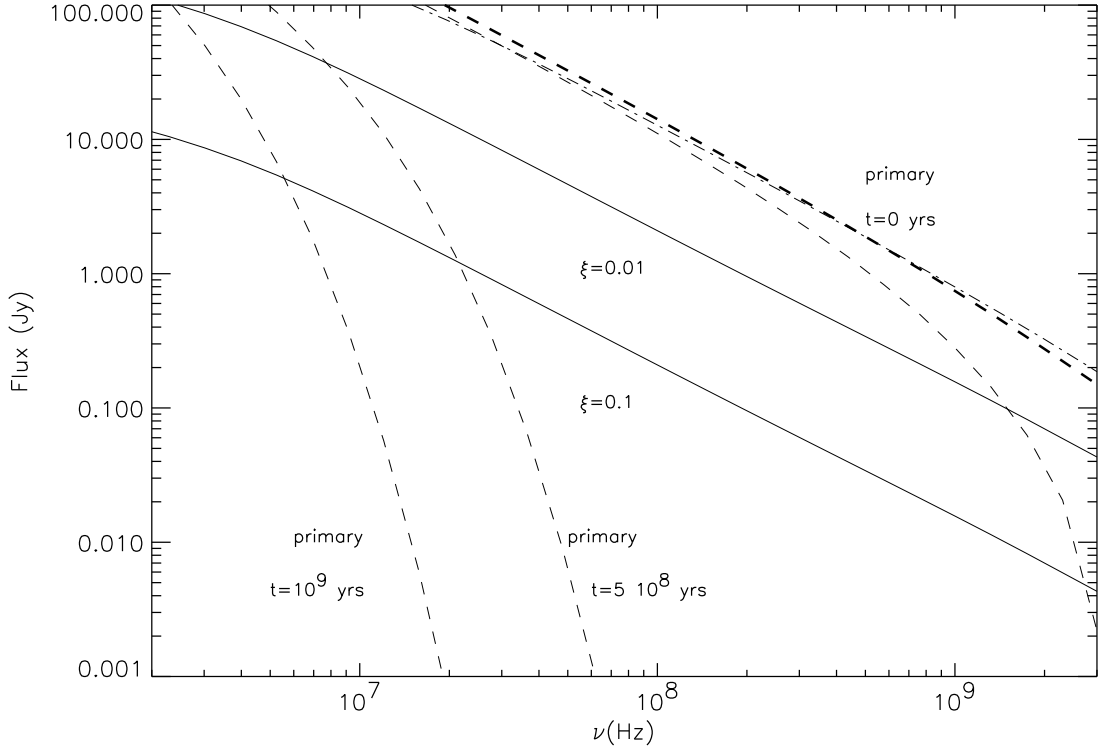,width=15.cm}}
  \caption{\em {Radio fluxes from a cluster at $100$ Mpc distance. The three
dashed thin lines refer to primary electrons at different times after the 
merger, as indicated. The solid lines are the contributions of secondary
electrons for two values of $\xi$. The dashed thick line is the radio flux 
from primary electrons at $t=0$ in case of Bohm diffusion. The dash-dotted line
is the synchrotron emission from primary electrons at $t=0$ in the case 
$B=1\mu G$.
}}
 \end{center}
\end{figure}

\begin{figure}[thb]
 \begin{center}
  \mbox{\epsfig{file=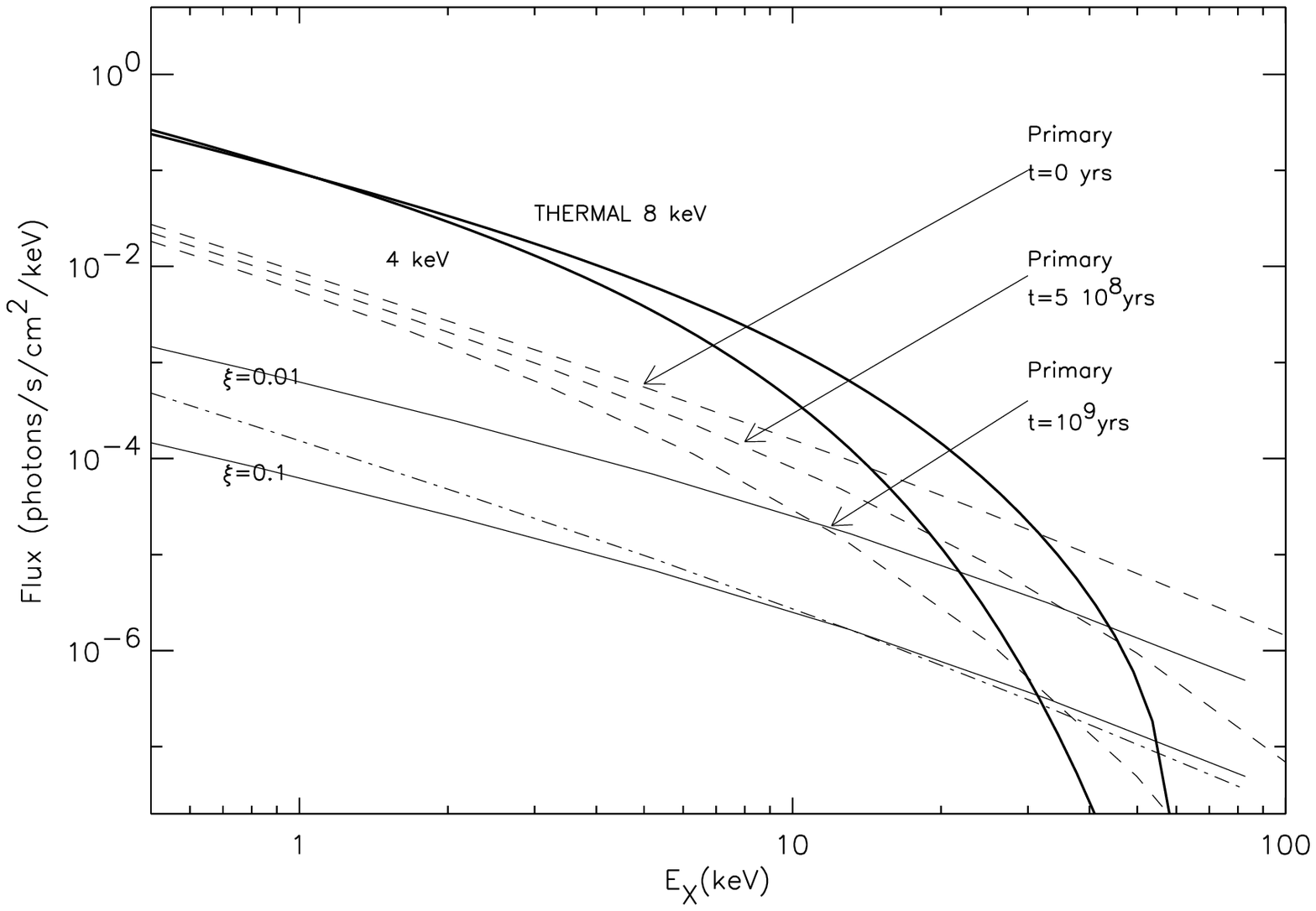,width=15.cm}}
  \caption{\em {X-ray fluxes from a cluster at $100$ Mpc distance. The thick
solid lines are the thermal bremsstrahlung emission of the thermal gas at
a temperature of 8 KeV and 4 keV. The other curves are the ICS constribution 
of non-thermal electrons (labelled as in Fig. 1). The dash-dotted line 
refers to the case $B=1\mu G$, to emphasize the fact that X-ray fluxes
are suppressed for high magnetic fields.
}}
 \end{center}
\end{figure}

It is immediately clear that 
already after $5\times 10^8$ yrs the radio emission is limited to frequencies
less than $\sim 30$ MHz, not easily accessible to observations, and after 
$10^9$ yrs, the emission is negligible above $\sim 5$ MHz. There is no way that
primary electrons can produce a persistent radio emission up to a few
GHz (as in the case of Coma) if the merger event ended more that $\sim 20$
million years ago. 
On the other hand the radio emission due to secondary electrons is irreducible.
It persists in the cluster even if the merger occurred at the beginning of the
cluster hystory and it is appreciable even for the conservative value 
$\xi=0.1$. The solid curves in Fig. 1 remain unchanged with time, after the
end of the merger event.

In Fig. 2 a similar situation is illustrated for the range of energies between
$100$ eV and $100$ keV. The thick solid curves represent the bremsstrahlung
emission of a gas of thermal electrons at $8$ keV and $4$ keV temperature,
as indicated. The 
dashed lines represent the contribution due to ICS of the primary
electrons off the photons of the cosmic microwave background, for
three different times after the end of the merger. 
It is possible to see that the X-ray excess at $E_X>30$ keV (10 keV)
is basically negligible for times larger than $\sim 10^9$ years after
the merger for a temperature of 8 keV (4 keV). 
Clearly the non-thermal fluxes are more prominent in the lower temperature
case.

As for Fig. 1, the dash-dotted line represents the X-ray flux from primary
electrons for a magnetic field of $1\mu G$, immediately after the end of
the merger event. We can clearly see that the X-ray
flux in this case is about $\sim 60$ times lower than for the case 
$B=0.15\mu G$.

The ICS contribution due to secondary electrons is not time
dependent and is therefore present even if the merger occurred at the 
epoch of the cluster formation. If the primary electrons are responsible
for the hard X-ray excess, then we have to be in a quite narrow window of
times after the merger, otherwise the contribution fades away due to 
electron energy losses. The possible small excess at energies below $\sim
1$ keV is not appreciably time dependent. This is due to the fact that the
cooling time for the low energy electrons is relatively long. 
One comment that is probably worth making is the following: if one compares 
the results in Figs. 1 and 2, it is possible to note that, for instance
$5\times 10^8$ yrs after the end of the merger, the radio flux in the region
accessible to observations is basically absent, while there is an appreciable 
X-ray flux below $\sim 1$ keV and above $20-30$ keV. This implies that the
detection of an X-ray excess without a corresponding radio excess does not
automatically imply an upper limit on the magnetic field. This should be 
taken into account when an upper limit on the magnetic field of a cluster
is derived from these considerations \cite{limit}, by imposing that the
radio emission is too low to be detected.

Any model for the soft or hard X-ray emission must face the gamma ray limit. 
The gamma ray emission from a cluster at $100$ Mpc is plotted in Fig. 3. The
thick solid lines represent the gamma ray fluxes due to pion decay in $pp$ 
scattering for $\xi=0.1$ (curve Pion01) and $\xi=0.01$ (curve Pion001). 
The three dashed lines represent
the gamma ray flux due to bremsstrahlung of the primary electrons at the three
different times used in Figs. 1 and 2. The dash-dot-dot-dotted lines are
the fluxes of gamma rays due to bremsstrahlung emission of the secondary 
electrons for $\xi=0.1$ (curve BS01) and $\xi=0.01$ (curve BS001). 
The thin solid lines are the gamma ray fluxes due to
ICS of the secondary electrons. The upper curve is for $\xi=0.01$ (ICS001) 
and the lower curve is for $\xi=0.1$ (ICS01).
All these cases are referred to eq. (\ref{eq:dif1}) as diffusion coefficient
but are not appreciably affected by other choices. 
For the ICS of primary electrons the situation is different: in the
diffusion scenario adopted here [eq. (\ref{eq:dif1})] the maximum energy 
of the primary electrons is
too low to generate appreciable gamma ray fluxes in the GeV region. However,
for a bohm diffusion [eq. (\ref{eq:dif2})] the maximum energy becomes large 
enough and the correspondent gamma ray flux is plotted in fig. 3 as a dotted
line. 

\begin{figure}[thb]
 \begin{center}
  \mbox{\epsfig{file=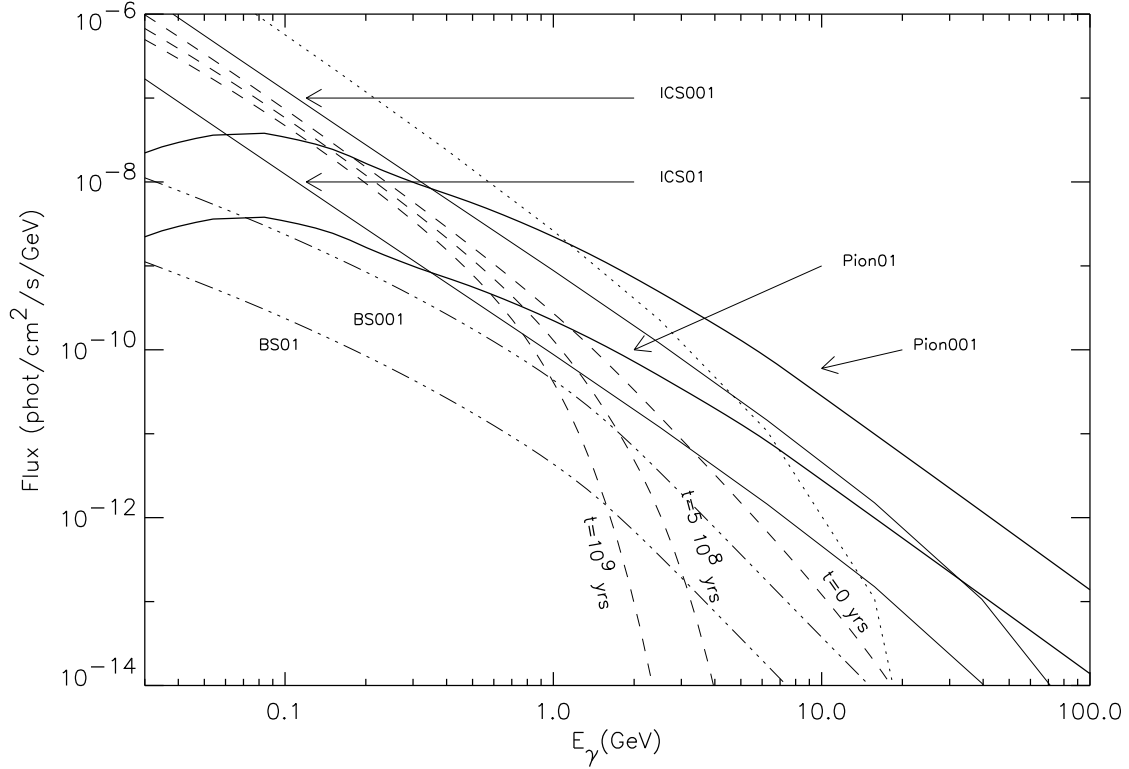,width=15.cm}}
  \caption{\em {Gamma ray fluxes from a cluster at $100$ Mpc distance. The
dashed lines are the bremsstrahlung contribution from primary electrons at
different times after the merger, the thick solid lines represent the gamma 
rays from pion decay for $\xi=0.1$ (curve Pion01) and $\xi=0.01$ 
(curve Pion001), the dash-dot-dot-dotted lines represent the bremsstrahlung 
emission from secondary electrons for $\xi=0.1$ (curve BS01) and $\xi=0.01$
(curve BS001). The thin solid lines represent the ICS contribution from 
secondary electrons for $\xi=0.1$ (curve ICS01) and $\xi=0.01$ (ICS001).
The dotted line is the ICS gamma ray flux from primary electrons at $t=0$
for a Bohm diffusion coefficient. At later times this contribution rapidly
fades away.
}}
 \end{center}
\end{figure}

It is easy to recognize the strength of the gamma ray limit. 
Note that even if the total energy in primary electrons is only
$\sim 1-3\%$ of the merger energy input, the gamma ray fluxes are 
comparable with the EGRET sensitivity (for a distance of $100$ Mpc).
Moreover, if the maximum energy of the primary electrons is larger than
$\sim 200$ GeV, a large gamma ray flux is produced by ICS.

Another feature which is evident from Fig. 3 is that while the gamma ray 
fluxes at energies above $\sim 1$ GeV contributed by primary electrons
fastly fade away, the contribution of gamma rays from pion decay and ICS 
of secondary electrons remains
important. Actually, as proposed in \cite{blasi99}, the flux of gamma rays
above $10-100$ GeV can be used as a direct 
tool to weigh the cosmic ray energy content of clusters. This is due to the
fact that {\it a)} protons do not lose energy appreciably, {\it b)} the 
spectrum of the gamma rays from pion decay follows the same spectrum of the
parent protons at high energy, {\it c)} the spectrum of electrons and 
therefore also the spectrum of the bremsstrahlung emission, is steepened by
the electron energy losses, so that their contribution at high energy is
negligible.

\section{The case of the Coma cluster}

As an application of the calculations illustrated above, we present the 
case of the Coma cluster, where excesses in UV and X-rays are detected, 
as well as a radio halo extended up to frequencies of a few GHz. 
In the merger model
this situation can be realized if the merger has just occurred, and indeed
there are some indications that this might be the case for Coma.

If Fig. 4 we plotted the results of our calculations for the radio emission,
on top of the data points obtained in \cite{feretti}. 
The (solid) line that fits the data
is the contribution of the primary electrons with an injection rate
correspondent to $\sim 2\%$ of the total luminosity of the merger 
$L_{merger}$ when the magnetic field is $0.15\mu G$. The dash-dotted line
represents the case $B=1\mu G$, in which only $0.03\%$ of $L_{merger}$ is
needed to explain the radio observations. 
$L_{merger}$
has been estimated as the ratio of the total thermal energy in the cluster
within $1$ Mpc from the center, and the merger duration ($\sim 10^9$ yrs). 
The number we obtained can be considered as an upper limit, since the gas
was presumably already hot before the merger.
The diffusion coefficient was chosen as in eq. (\ref{eq:dif1}) with 
$v_8\approx 4.8$ and $L_{20}\approx 10$ (200 kpc, correspondent to the
typical distance between two galaxies in Coma). These values were chosen only
in order to obtain a maximum energy for the primary electrons large enough to
generate GHz radio waves. In the case of Bohm diffusion this problem does
not appear, but there is no cutoff at GHz frequencies (see fig. 1).

\begin{figure}[thb]
 \begin{center}
  \mbox{\epsfig{file=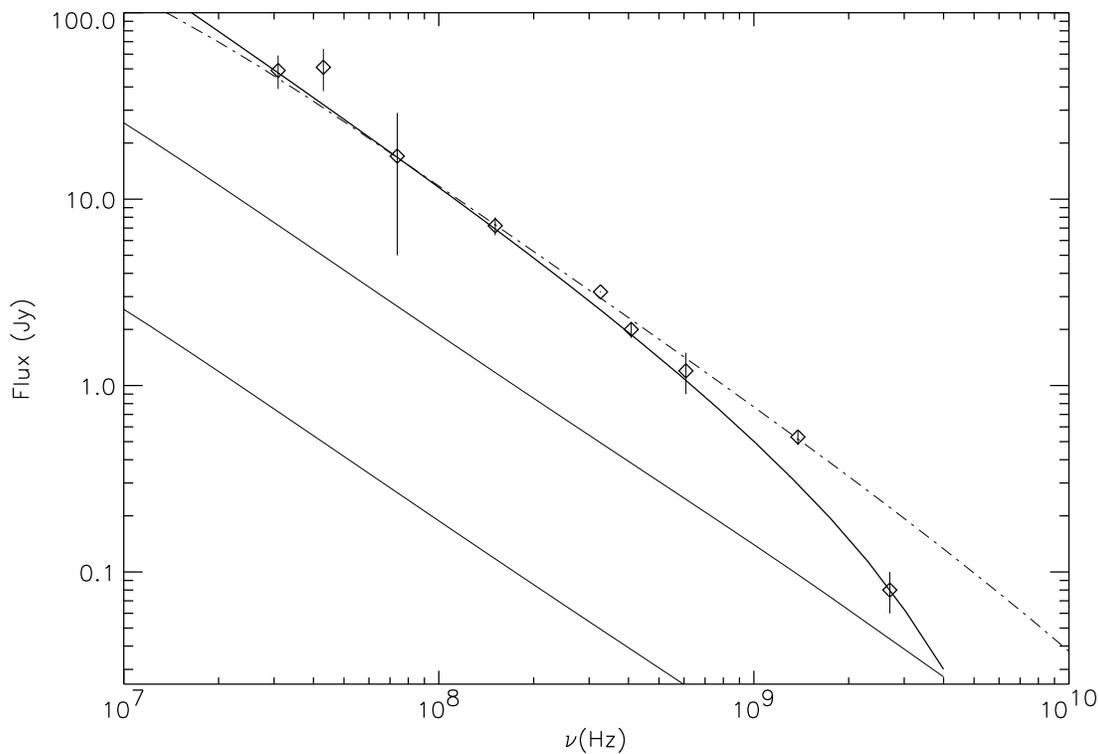,width=15.cm}}
  \caption{\em {Radio emission from the Coma cluster. The data points from
\cite{feretti} are well fitted by the primary electron synchrotron
emission for $B=0.15\mu G$. The other two solid curves represent the 
contribution of secondary electrons for $\xi=0.1$ (lower curve) and 
$\xi=0.01$ (upper curve), again for $B=0.15\mu G$. The dash-dotted line
is the synchrotron flux from primary electrons for $B=1\mu G$.
}}
 \end{center}
\end{figure}

The two bottom lines in Fig. 4 are the contribution of the secondary electrons
for $\xi=0.1$ (correspondent to $\sim 5\%$ of $L_{merger}$) and $\xi=0.01$
(corresponding to $\sim 46\%$ of $L_{merger}$). 

The corresponding ICS contributions to UV and soft/hard X-rays are plotted in
Fig. 5, together with the thermal contribution from a gas at the temperature
of Coma (thick solid line). The data points are from Beppo-SAX, while the 
dark region is an estimate of the soft X-ray/UV flux \cite{lab}.
The thin solid line is the ICS contribution of primary electrons
for $B=0.15\mu G$: both the UV and hard X-rays are quite well described 
by this curve.
The two dashed lines are the ICS fluxes of secondary electrons for
$\xi=0.1$ (lower curve) and $\xi=0.01$ (upper curve). 
The dash-dotted line is the ICS flux from primary electrons for
$B=1\mu G$. It is evident
that in this case the relativistic electrons give a negligible contribution to
the X-ray flux, therefore, for the large fields found by Faraday
rotation measurements, ICS cannot be invoked as an explanation of the hard
X-ray excess. For fields even higher than $1\mu G$ clearly the problem 
becomes more evident.

\begin{figure}[thb]
 \begin{center}
  \mbox{\epsfig{file=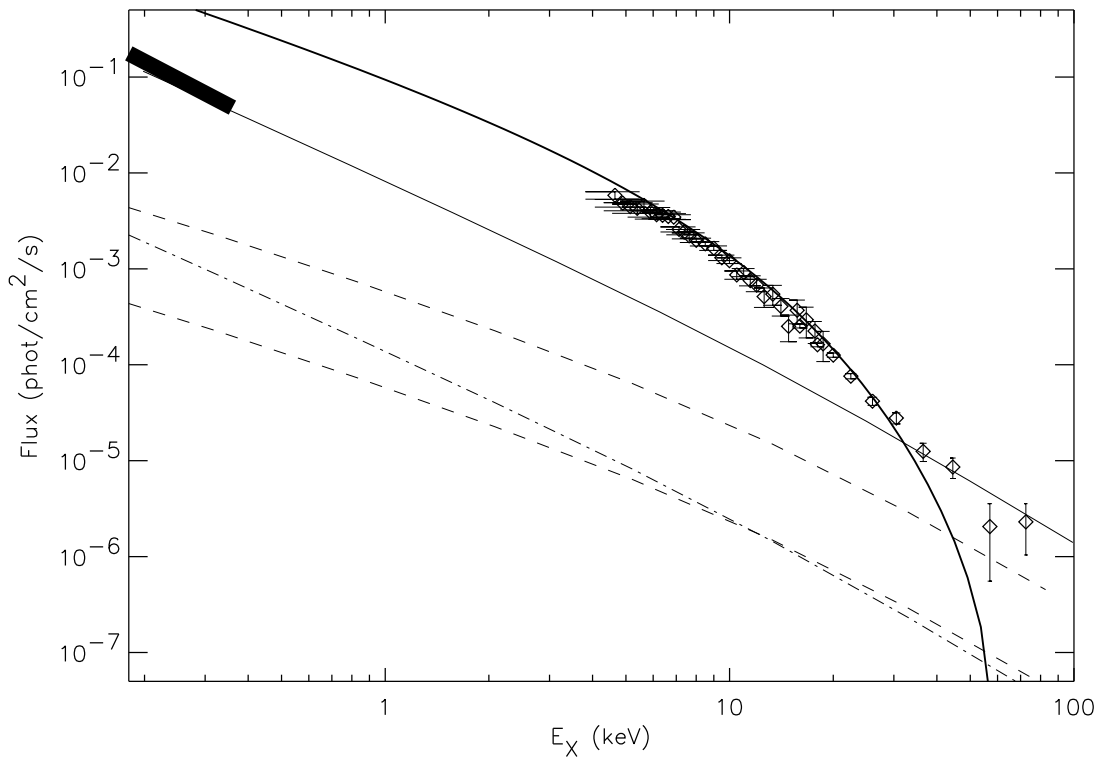,width=15.cm}}
  \caption{\em {X-ray emission from the Coma cluster. The data points are from
\cite{fusco}. The solid thick line is the bremsstrahlung emission of thermal
gas, the solid thin line is the ICS flux from primary electrons, and
the two dashed lines are the ICS fluxes from secondary electrons for the two
usual values of $\xi$ and $B=0.15\mu G$. 
The dark region is an estimate of the low energy X-ray excess from \cite{lab}.
The dash-dotted line is the X-ray flux expected from primary electrons
for $B=1\mu G$, too small by a factor $\sim 60$.
}}
 \end{center}
\end{figure}

Is the low magnetic field situation compatible with the gamma ray bound? 
In Fig. 6 we plotted the 
gamma ray flux from Coma, separating the contribution of the bremsstrahlung 
from the primary electrons (thick line) and secondary electrons (dash-dotted
lines), the pion decay for the usual 
two values of the parameter $\xi$ (thin solid lines) and 
the ICS of the secondary electrons (dashed lines). The dotted line
represents the ICS flux from primary electrons when a Bohm diffusion is
used. In the other case, for the reasons explained above there is no
gamma ray flux due to ICS of primary electrons. 
The gamma ray 
flux at $\sim 100$ MeV exceeds the EGRET upper limit \cite{egret}
by a factor $\sim 2-4$ for the diffusion coefficient in eq. (\ref{eq:dif1})
and by a factor $\sim 15$ for the Bohm diffusion.

\begin{figure}[thb]
 \begin{center}
  \mbox{\epsfig{file=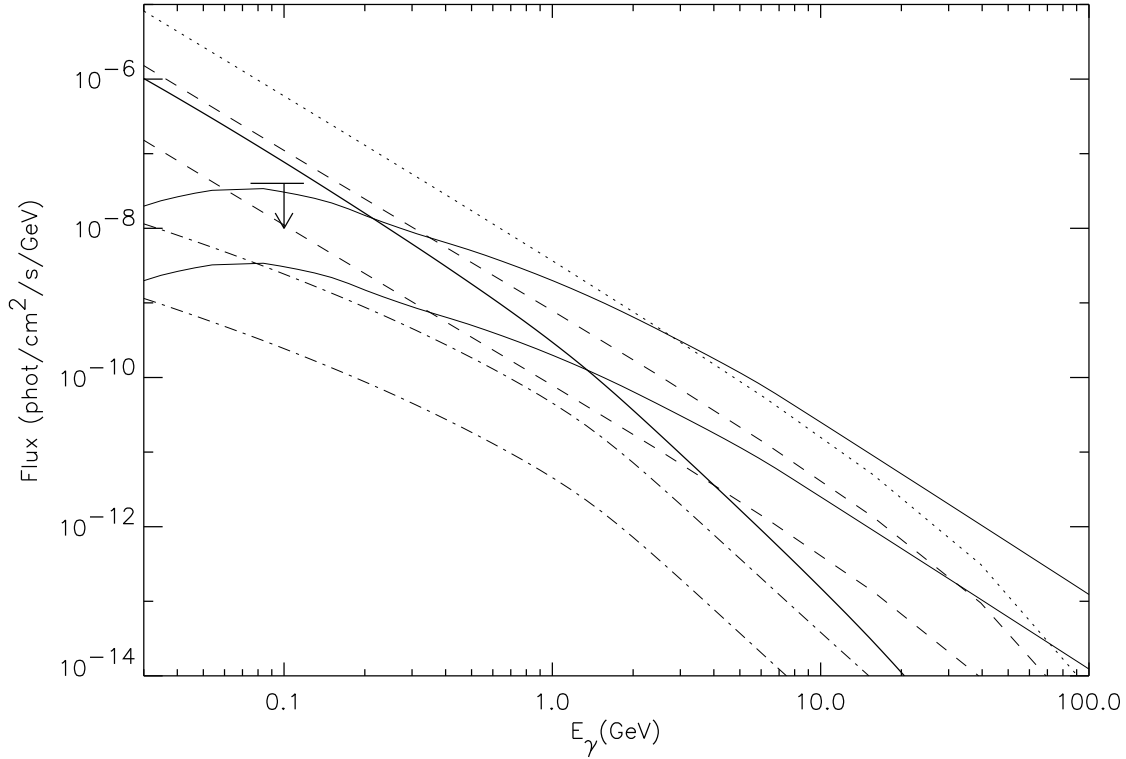,width=15.cm}}
  \caption{\em {Gamma ray fluxes from the Coma cluster. The EGRET upper 
limit is from \cite{egret}. The solid thin lines are the result of pion
production for $\xi=0.1$ (lower curve) and $\xi=0.01$ (upper curve). The
solid thick line is the bremsstrahlung contribution of primary electrons. The
dashed lines are the ICS fluxes from secondary electrons for 
$\xi=0.1$ (lower curve) and $\xi=0.01$ (upper curve). The dotted line is the
ICS gamma ray flux from primary electrons in the case of Bohm diffusion.
}}
 \end{center}
\end{figure}

The estimate of the gamma ray excess at $100$ MeV may be affected by
the assumptions on the spatial distribution of the injected cosmic rays, 
but it is difficult to envision how to reduce the predicted flux in
such a way that the results become compatible with the EGRET limit.
It is worth to stress again the role of current and
future gamma ray detectors to clarify the origin of the non-thermal radiation
from clusters of galaxies, as already discussed in \cite{blasi99}: if the
flux of gamma rays from Coma is just below the EGRET upper limit, 
additional efforts for its detection would
certainly help in understanding the source of the radiation at other
wavelengths.

\section{Discussion and Conclusions}

The role of mergers of clusters of galaxies for the heating of
the intracluster gas has already been recognized through observations 
as well as through numerical simulations. The heating mainly occurs 
due to the formation of collisionless shocks, whose presence shows up
in simulations and in observations through the appearance of hot rims.
Being the typical compression factor of these shocks in the range 
$r\sim 3-4$, they may be of some relevance for the acceleration of 
a fraction of the thermal gas to ultra-relativistic energies, by first
order Fermi acceleration. 

We studied here the non-thermal radiation generated by 
electrons and protons energized at a typical merger shock in a generic
cluster. The aim of this calculation was not to consider a specific 
case, but rather to draw some general conclusions about the role of mergers
as sources of the non-thermal radiation recently detected from several 
clusters of galaxies.

The main problem with reaching this goal is that mergers are short
events on cosmological time scales, but relatively long duration events
on the scale of energy losses of electrons in the CMB bath. Therefore,
whether a merger can induce or not an appreciable flux 
of non-thermal radiation 
through ICS and synchrotron emission of high energy electrons depends on
how long ago the cluster merger occurred, in terms of acceleration of new
particles. This is not true for the secondary electron component, generated
through the decay of charged pions in $pp$ interactions. Due to the confinement
of cosmic ray protons in the cluster \cite{bbp}, the contribution to the
non-thermal radiation, due to secondary electrons is basically time independent
after the end of the merger event.

A possible summary of the results obtained here is as follows:

{\it i)} As shown in Fig. 1, the radio emission due to primary electrons 
is dramatically time dependent. If the injection of newly accelerated 
particles ends (the merger is over), after $5\times 10^8$ yrs the radio
emission above $\sim 30$ MHz has already disappeared. The radio emission 
due to secondary electrons is a function of the ratio of protons to electrons, 
but, as illustrated in Fig. 1, this contribution is appreciable for a cluster
at a typical distance of $100$ Mpc. As stated before, this emission is
time independent, therefore, if a diffuse radio emission is detected in a 
cluster where there is no current evidence of a very recent merger, this
would indicate the presence of
a substantial relativistic hadronic component in the intracluster medium.

{\it ii)} As far as the X-ray emission is concerned, it is quite well 
proven that the bulk of the X-ray radiation from clusters of galaxies is 
due to bremsstrahlung of thermal electrons at a temperature of $10^7-10^8$ K.
The soft and hard X-ray and the extreme UV excesses recently detected from 
some clusters need further investigation: for the UV excess, there seem
to be different tendencies, from a thermal interpretation 
\cite{bonamente01,bonamente02} to a ICS
interpretation \cite{relic,ensslin}. In the hard X-ray region the evidence for 
an excess seems to be more solid. 
In Fig. 2 we plotted the ICS contribution in the region of interest, by dashed
lines (for $B=0.15\mu G$). Although after $5\times 10^8$ yrs the X-ray flux 
above $\sim 30$ keV is considerably reduced, it seems still 
appreciable, while, as stated above, the
correspondent radio flux above $30$ MHz has faded away. This means that there
is a time window where there might be X-ray activity and no radio flux. In this
case it is not possible to extract upper limits on the magnetic field from 
the absence of radio emission. 
The dash-dotted line refers to the case $B=1\mu G$ and makes clear that 
large magnetic fields considerably decrease the amount of non-thermal
X-ray emission.

The X-ray flux generated by ICS of secondary electrons may be appreciable, 
and, as for the radio flux, it is time independent, so that it is still 
present even when the whole contribution from primary electrons has
disappeared. 
Fig. 2 also makes clear that it is more promising to look for non-thermal
excesses in lower temperature clusters, where the contribution due to thermal
bremsstrahlung disappears at lower energies, leaving the hard X-ray fluxes
to dominate.

{\it iii)} What is the overall energy requirement in order to have reasonably
large non-thermal fluxes and how large can the magnetic field be? 
We found that a few percent of $L_{merger}$ in the
form of accelerated electrons can explain the observations if the merger is
very recent (see for instance the case of Coma in section 6). 
This issue is directly related to the strength of the magnetic field 
in the ICM. This is an open issue: it is not easy to envision a simple
reason why Faraday rotation measurements (that always provide a result
smaller than the ``real'' field) actually give values higher or even 
much higher \cite{eilek,tracy} than the field required 
to explain the multiwavelength observations of clusters.
At present there is no theoretical argument that may help in 
estimating the value of the magnetic field in clusters, since its
origin is a complete mystery. The intracluster field  might be 
cosmological \cite{olinto}, or
might as well be generated in shock waves during the process of formation
of large scale structures \cite{kang} or expelled by radio galaxies 
\cite{daly,okoye} or by galactic winds \cite{KLH,VA99}. In any case the 
theoretical input on this issue is not significant and the main source
of information on the magnetic field is either through indirect measurements
(radio plus X-rays) or by measurements of the Faraday rotation, and as we
stressed above, these two methods give results which do no seems compatible
with each other.

The need for small magnetic fields is common to all models based on ICS
(for the hard X-rays) and synchrotron (for the radio emission) and is not 
limited to cluster merger scenarios. This seems to us as a strong argument
against this class of models, that should stimulate the search for alternative
explanations. The problem is evident from Figs. 1 and 2: the dash-dotted lines
are the non-thermal fluxes for the case $B=1\mu G$; the radio emission can 
be at the same level as in the case $B=0.15\mu G$, with about $60$ times less
electrons, but the X-ray fluxes are correspondingly lower by a factor $60$.

Is there any way to reconcile the results of Faraday rotation measurements with
the magnetic field required by non-thermal observations? One possibility is
that regions of very high magnetic field are separated by regions where the
magnetic field is lower. In these conditions the Faraday rotation might be 
affected mainly by the strongly magnetized regions while the hard X-rays
would come from the low field regions. This scenario might imply a 
phase separation in the intracluster medium, which is actually observed in some
cases (see for instance \cite{phase}). However such a situation does not seem
to be happening in clusters like Coma or other clusters that show 
non-thermal activity and appear relaxed. Moreover, the flux of hard X-rays 
is basically independent of the magnetic field (it is the normalization between
X-rays and radio emission that implies low magnetic fields), therefore the 
large gamma ray fluxes studied in section 6 would not be affected by this
argument, and in the case of the Coma cluster the predicted gamma ray fluxes
would still be in excess of the EGRET upper limit.

For this reason gamma ray observations can be considered as a more solid 
constraint on the total amount of energy that can be injected in the 
cluster (therefore indirectly also on the strength of the average
magnetic field). Gamma ray emission in a
cluster is produced mainly by a) bremsstrahlung emission of primary and
secondary electrons, b) ICS of primary and secondary electrons, 
and c) decay
of neutral pions generated in $pp$ interactions. The results are plotted
in Fig. 3: the bremsstrahlung of primary electrons fades away quite rapidly,
but it remains important below a few hundreds MeV, accessible to EGRET. The
bremsstrahlung of the secondary electrons is usually quite small. A very
important contribution to the gamma ray emission is provided by ICS of the
secondary electrons. It is worth recalling again that this flux is not
time dependent, therefore it does not fade away with time, even if the merger
occurred in the distant past of the cluster. The same is true for the gamma ray
emission due to pion decay. The ICS contribution of primary electrons 
depends on the maximum energy of the electrons, which is in turn affected by
the choice of the diffusion coefficient.

To make the prediction more quantitative, we applied our calculations to the
case of the Coma cluster. For a merger that just ended, the radio flux 
can be fitted quite well by the synchrotron emission of primary electrons.
The results are plotted in Fig. 4. The cut off at a few GHz is here the result
of a cutoff in the spectrum of the accelerated electrons and it could be 
absent for a different model of diffusion (the existence of the cutoff in the
observations has been questioned anyway in \cite{deiss}). The value of the 
magnetic field required to fit simultaneously the radio and hard X-ray 
emission is $B=0.15\mu G$ for an emission volume of $1$ Mpc. With these
parameters the soft X-ray and UV excesses are also of the correct order
of magnitude. A larger field of $B=1\mu G$ is used to obtain the dash-dotted
line, corresponding to a rate of injection of electrons $\sim 60$ times
lower than in the previous case. For $B=1\mu G$ the X-ray flux by ICS 
is $\sim 60$ times too small to explain the BeppoSAX observations (Fig. 5). 

What is the role of protons in the Coma cluster? With the 
parameters used above, the contribution of secondary electrons to both radio
and hard X-ray emission is small, but for $\xi=0.01$ the EGRET upper limit to
the gamma ray flux above $100$ MeV is saturated. Indeed this limit is
exceeded by the bremsstrahlung of primary electrons added to the ICS
emission of secondary electrons (Fig. 6) and possibly of primary electrons
(if their maximum energy is large enough).

What happens if the merger occurred more than, say, $10^9$ yrs ago?
In this case, as discussed above, the contribution of primary electrons has
bocome negligible by now, therefore the secondary electrons must be 
responsible. From Figs. 4-6 we see that this implies a
large energy injection in cosmic ray protons, resulting in a flux 
of gamma radiation with energy larger than 100 MeV which is in excess of the
EGRET limit. 
Even if the merger occurred very recently the gamma ray fluxes are close to or
even in excess of the EGRET upper limit.

From the discussion above, it is clear that the interpretation of the
non-thermal multiwavelength observations from clusters (and in particular from
Coma) based on synchrotron and ICS of relativistic electrons has several severe
problems. Therefore it is natural to look for some alternative explanations.
We address this issue briefly: is it possible that mergers play a
role in the production of non-thermal radiation and, at the same time, that 
intracluster magnetic fields are at the level measured through 
Faraday rotation
keeping the gamma ray fluxes below the EGRET upper limit?
As we showed, if ICS and synchrotron emissions are used as mechanisms for the
production of the 
non-thermal radiation, then the answer is likely to be negative.
However, in \cite{blasi00} a viable alternative was presented: during the
merger the perturbation of the magnetic field results in the production of 
waves that can resonate with electrons on the tail of the thermal electron
distribution. Since waves and electrons are coupled non-linearly, part
of the energy released in the tail of the Maxwell-Boltzmann distribution
ends up in the low energy part of the electron spectrum and is rapidly 
redistributed in the form of thermal energy of the plasma. In other words,
it was shown that the thermalization process in the presence
of a magnetic field and a perturbation (e.g. the merger) is not a simple
process, and while the bulk of the thermal electrons is ``just'' heated up, the
fraction of electrons more weakly bound to the thermal bath can acquire a
non-Maxwellian distribution. If this happens, then the hard X-rays can be the
result of bremsstrahlung emission of the non-thermal tail 
\cite{blasi00,ensslin,dogiel}, without affecting at all the radio emission, 
that can
be generated through synchrotron radiation of a population of electrons
radiating in a magnetic field consistent with Faraday rotations. The gamma 
ray fluxes would also be insignificant. This
scenario has two nice features: 1) it provides a unified picture for the
heating process and for the energization of low energy electrons to non-thermal
energies; 2) it explains the thermal and non-thermal X-ray emission in 
terms of the same process (bremsstrahlung emission).\par\noindent
In \cite{bos} it was proposed a clear way of testing this scenario, by
using precision measurements of the Sunyaev-Zeldovich effect from clusters
of galaxies showing non-thermal activity.

{\bf Aknowledgments} I am grateful to C. Sarazin and P.L. Biermann 
for discussions. I am also grateful to R. Fusco-Femiano for kindly 
providing the BeppoSAX data of the X-ray emission from the Coma cluster
and to the anonymous referee for useful comments.
This work was supported by the DOE and the NASA grant NAG 5-7092 
at Fermilab.

\end{document}